\documentclass[12pt,preprint]{emulateapj}

\shorttitle{3D MHD modeling of propagating disturbances in coronal loops}
\shortauthors{Wang et al.}

\begin{document}

\title{Three-dimensional MHD modeling of propagating disturbances in fan-like coronal loops}

\author{Tongjiang Wang\altaffilmark{1,2}, Leon Ofman\altaffilmark{1,2,3}, and 
Joseph M. Davila\altaffilmark{2} \\
Recevied: 17 May 2013;~~~ Accepted: 12 August 2013}
\altaffiltext{1}{Department of Physics, Catholic University of America,
   620 Michigan Avenue NE, Washington, DC 20064, USA; tongjiang.wang@nasa.gov}
\altaffiltext{2}{NASA Goddard Space Flight Center, Code 671, Greenbelt, MD 20770, USA}
\altaffiltext{3}{Visiting Associate Professor, Tel Aviv University, Israel}

\begin{abstract}
Quasi-periodic propagating intensity disturbances (PDs) have been observed in large 
coronal loops in EUV images over a decade, and are widely accepted to be slow magnetosonic 
waves. However, spectroscopic observations from Hinode/EIS revealed their association 
with persistent coronal upflows, making this interpretation debatable. Motivated by 
the scenario that the coronal upflows could be cumulative result of numerous individual flow 
pulses generated by sporadic heating events (nanoflares) at the loop base, we construct 
a velocity driver with repetitive tiny pulses, whose energy frequency distribution 
follows the flare power-law scaling. We then perform 3D MHD modeling of an idealized bipolar 
active region by applying this broadband velocity driver at the 
footpoints of large coronal loops which appear open in the computational domain. Our model 
successfully reproduces the PDs with similar features as the observed, and 
shows that any upflow pulses inevitably excite slow magnetosonic 
wave disturbances propagating along the loop. We find that the generated PDs 
are dominated by the wave signature as their propagation speeds are consistent with 
the wave speed in the presence of flows, and the injected flows rapidly decelerate with height. 
Our simulation results suggest that the observed PDs and associated persistent upflows 
may be produced by small-scale impulsive heating events (nanoflares) at the loop 
base, and that the flows and waves may both contribute to the PDs at lower heights.

\end{abstract}


\section{Introduction}
 Quasi-periodic propagating disturbances (PDs) have been found in large coronal loops 
\citep[e.g.][]{nig99, ber99} since 
the launch of SOHO and TRACE. The PDs usually propagate at a speed on the order of
100 km~s$^{-1}$ which is close to the sound speed in the warm ($\sim$1 MK) corona,
leading to their interpretation as slow magnetosonic waves (see, \citealp{ofm99, nak00},
and a review by \citealp{dem09}). Under this assumption, the observed amplitude 
decay was explained in terms of compressive viscosity \citep{ofm00} or 
thermal conduction \citep{nak00, dem04}, and the quasi-periodic nature of these waves 
has been attributed to the leakage of $p$-modes \citep{dep05}.  

Alternatively, the PDs were also interpreted as recurring flows (or jets) because
of their association with high-speed upflows in footpoint regions of AR loops found by
Hinode/EIS \citep[e.g.][]{sak07, har08, uga11}. 
Recent finding that quasi-periodic variations in intensity, Doppler velocity, line width, 
and line asymmetry are correlated in the upflow regions supported 
this interpretation \citep{dep10, tian11a, tian11b}. 
 
Recently, using the 3D MHD model, \cite{ofm12} studied the excitation 
and propagation of slow magnetosonic waves in hot ($\sim$6 MK) closed loops driven
by steady or periodic upflows. Here we expand their model to simulate the PDs in 
large warm loops using a broadband flow driver in order to understand 
the physics of the observed PDs. 

\begin{figure*}
\epsscale{0.9}
\plotone{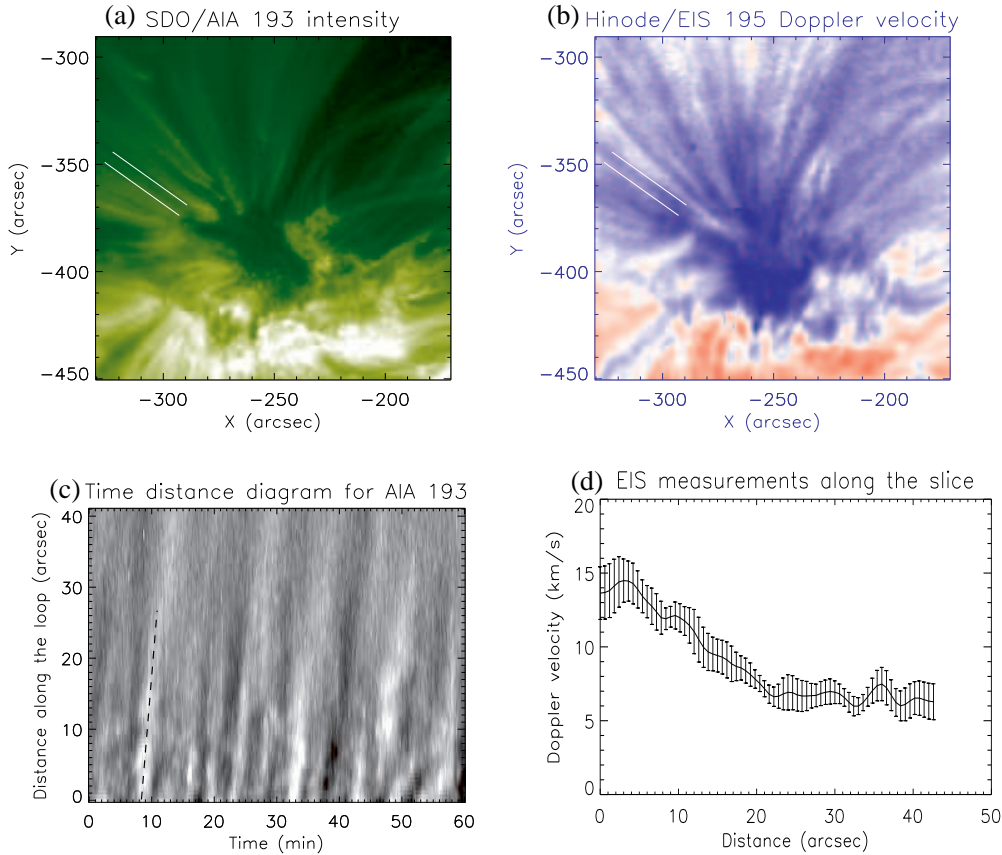}
\caption{ \label{fgobs} Observations of propagating intensity disturbances and outflows 
on 2010 September 16 in AR 11106. (a) The image of SDO/AIA 193 \AA\ at 11:57 UT. 
(b) The Doppler shift map of Hinode/EIS Fe\,{\sc{xii}} 195.12 \AA\ scanned from 
10:38 to 11:57 UT. (c) A time-distance diagram of AIA 193 \AA\ detrended intensity for
a cut (shown with two parallel lines in (a)), averaged over a width of 11 pixels. 
The inclined dashed line indicates one propagating disturbance with a traveling 
speed of about 128 km~s$^{-1}$. (d) The Doppler velocity profile of EIS Fe\,{\sc{xii}}
along the cut (shown in (b)).}
\end{figure*}

\section{Observational motivation}
There has been plenty of evidence supporting that the PDs in coronal loops above sunspots
are propagating slow magnetosonic waves. For example, the oscillations there have a period 
of about 3 minutes, same as the well-known umbral oscillations \citep{dem02}. The
multi-wavelength observations indicate that the waves propagate from the photosphere through
the chromosphere and transition region into the corona \citep[e.g.][]{bry02, mar03, jes12}.
 The propagation speeds of PDs are temperature dependent
\citep{kid12, uri13}. The PDs propagate outward (e.g. in AIA 171) in the circular or arc shape, 
similar to those observed in the chromosphere \citep{lia11,lia12}.
In contrast, the PDs in non-sunspot coronal loops appear to be different in some features. 
Their periods cover a wide range of about 3$-$30 min 
\citep{dem02, ber99, mar09, wan09, kri12}. The origin of those oscillations 
with longer ($\ga$10 min) periods is hardly explained by
the leakage of global $p$-modes \citep{dep05}. In addition, they
are usually associated with distinct upflows 
\citep[e.g.][]{tian11b}, and the coherent scale of oscillations is relatively small
(on the order of a few Mm) \citep{mce06, kid12}.
These differences suggest that the PDs in non-sunspot loops may have 
a different driver from the PDs in sunspot loops. One possible source could be
recurrent small-scale impulsive events (nanoflares) at the coronal base
\citep{ofm12, tes13}.

Figure~\ref{fgobs} demonstrates a typical AR with upflows. The time-distance diagram
shows the PDs propagating at almost constant speed of about 130 km s$^{-1}$ for a cut
along fan-like loops in SDO/AIA 193 \AA\ (Figures~\ref{fgobs}(a) and (c)). 
The Doppler velocity, measured by single Gaussian fits to the EIS Fe\,{\sc{xii}} 
195.12 \AA\ line, shows a decrease from 15 km~s$^{-1}$ to 6 km~s$^{-1}$ along the 
same cut (Figures~\ref{fgobs}(b) and (d)). The purpose of our modeling is to understand
the origin of PDs and their physical nature (flows or wave signatures).

\section{Numerical model}
\begin{figure*}
\epsscale{1.0}
\plotone{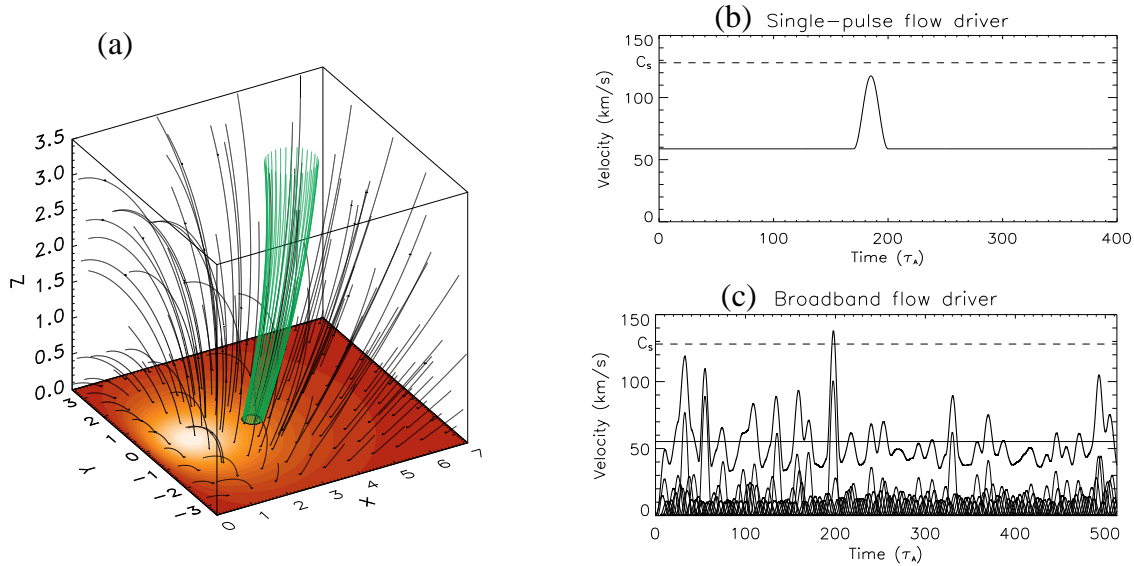}
\caption{ \label{fgmdl} (a) Initial magnetic field  used for the model AR,
which is taken as the half domain of a dipolar field. The intensity scale
shows the longitudinal field component magnitude at the base of the AR, and the black lines
show magnetic field lines. The green lines show a magnetic loop rooting on a flow region
at the bottom boundary. (b) Time profile of velocity for a single pulse flow driver. 
(c) A broadband upflow driver, which consists of 205 random pulses with kinetic energies 
following a powerlaw frequency distribution of the index $\alpha$=2. The horizontal
solid line indicates the resulting background level calculated by averaging the velocity 
over time.}
\end{figure*}

Here, we briefly describe the 3D MHD numerical model used in this study 
\citep[see,][for details]{ofm12}. The resistive 3D MHD equations are solved 
with gravity and isothermal energy equation on a Cartesian $258^3$ grid using the modified 
Lax-Wendroff method with fourth-order stabilization term \citep[e.g.,][]{ofm02}. The initial 
magnetic field is a dipole. To model an open-like loop with high spatial resolution, only
the half domain of the dipole is chosen for the 3D computation (Figure~\ref{fgmdl}), 
which is set as (0, 7) $\times$ ($-$3.5, 3.5) $\times$ (0, 3.5) in normalized distance units. 
The dipolar field is created by setting
two unit magnetic charges of the opposite polarities at the positions ($\pm$1, 0, $-$2).
The initial background density is given by the gravitationally stratified hydrostatic density
\begin{equation}
\rho = \rho_0 {\,\rm exp}\left(\frac{1/(10+z)-1/10}{H}\right) , \label{eqrh}
\end{equation}
where $H = 0.2 k_B T_0 R_s /(GM_s m_p )$ is the normalized gravitational scale height, 
$R_s$ is the solar radius, $k_B$ is Boltzmann's constant, $T_0$ is 
the temperature, $G$ is the universal gravitational constant, $M_s$ is the solar mass, 
and $m_p$ is the proton mass. The chosen and resulting normalization parameters are
listed in Table~\ref{tabpar}. \citet{van11} determined $\gamma$ (the adiabatic index) in loops 
and found that values are very close to unity expected in nearly isothermal plasma.

A localized impulsive flow injection along the magnetic field is introduced at the 
lower boundary of the model AR as 
\begin{equation}
{\bf V}=V_0(x, y, z=0, t){\bf B}/|B|, \label{eqvv}
\end{equation}
where 
\begin{equation}
V_0(x, y, z=0, t) =V_{A0} A_v(t){\,\rm exp}\left[-\left(\frac{r}{w_0}\right)^4\right],
\label{eqv0}
\end{equation}
with $r=[(x-x_0)^2+(y-y_0)^2]^{1/2}$.
The upflow is modeled with sharper-than-Gaussian cross-sectional profile, and
imposed at the lower boundary in a region with $r\le2w_0$. 
The parameters are $x_0$ = 3.0, $y_0$ = 0, and $w_0$ = 0.12.
We study two cases of upflow velocity: (1) a single pulse with constant background;
(2) multiple pulses with broadband frequency distribution of energy. In the first case, 
we take $A_v(t)$ in the form,
\begin{equation}
A_v(t)  =   \left\{
\begin{array}{ll}
 A_0 & \quad (0< t < t_1 \quad {\rm or} \quad t_2< t < t_{max}) \label{eqavt} \\ 
 A_0 + \frac{1}{2}A_0[1-{\rm cos}\frac{2\pi (t-t_1)}{P}] &  \quad (t_1\le t \le t_2) 
\end{array}\right.,
\end{equation}
where the parameters $P= t_2-t_1$=30, $t_1$=170, $t_2$=200, $t_{max}$=400, and $A_0$ = 0.01 
are set, so that the velocity driver has the maximum amplitude of 0.02, which is subsonic 
(Figure~\ref{fgmdl}(b)). In the second case, we construct a broadband driver with $N$ 
individual pulses of the same lifetime ($P$=15) and the same shape in the form
\begin{equation}
A_i(t)  = \frac{1}{2}A_{i0}[1-{\rm cos}\frac{2\pi (t-t_i)}{P}] \quad (t_i\le t \le t_i + P), ~
i=1,2,...,N 
\end{equation}
and assuming that they occur subsequently with a fixed rate ($\Delta{t}=t_{i+1}-t_i=P/6$).
We further assume the kinetic energy of flow pulses (defined as $E_i=A_{i0}^2$) follows 
a powerlaw frequency distribution ($df/dE \propto E^{-\alpha}$). For AR soft 
X-ray flares and microflares, it has been observed that the frequency distribution of 
their energy content has the index of about 1.8 \citep[e.g.][]{wan06}. Here we
randomly generate $N$ pulses for a powerlaw distribution with $\alpha=2$ and 
the velocity amplitudes in the range 10 to 110 km~s$^{-1}$. Figure~\ref{fgmdl}(c) shows 
the constructed velocity profile that includes about 26 visible peaks and a background 
(with a mean of 55$\pm$18 km~s$^{-1}$) over a time of 513 $\tau_A$.
The remaining boundary conditions are same as used in \citet{ofm12}.

In the low-$\beta$ condition of the nearly ideal coronal plasma, both the plasma motion 
and slow magnetosonic wave propagation are nearly along the magnetic field lines 
(i.e., the magnetic field is nearly unaffected by the flow). 
Thus we can determine whether the simulated PDs 
are the signature of flows or waves, by comparing their ``observed" paths
(traveling distances as a function of time) with those predicted.
Given the time-distance distribution ($V(s,t)$) of velocities along a loop, we 
calculate the flow path  by integrating the following quantity numerically,
\begin{equation}
s=\int_{t_0}^{t}V(s, t)dt, \label{eqflw} 
\end{equation}
and calculating the wave path by integrating
\begin{equation}
s=\int_{t_0}^{t}(V(s, t)+C_t(s))dt,  \label{eqwav} 
\end{equation}
with the initial condition $s$=0 at $t=t_0$, where $C_t(s)=(C_s^{-2}+V_A(s)^{-2})^{-1/2}$
is the tube speed for a straight cylinder \citep{rob84}.

\begin{figure*}
\epsscale{1.0}
\plotone{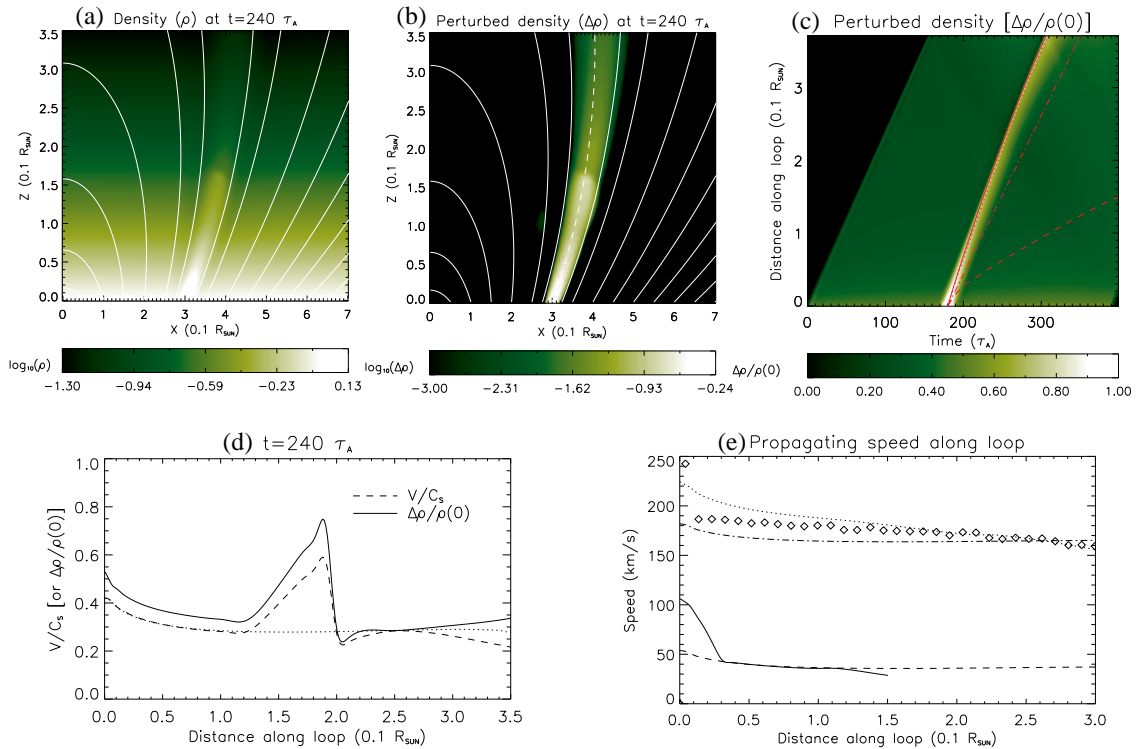}
\caption{ \label{fgsng} The modeling with a single pulse flow driver. (a) The density in 
the $xz$-plane at the center of the AR at t=240 $\tau_A$, and 
(b) Same as (a) but for the perturbed density. The overlaid white lines are magnetic
field lines in this plane. (c) Time-distance diagram of the perturbed density for a cut along 
the loop (dashed line in (b)). The solid and dashed lines are the predicted paths 
for the wave front and the flow, respectively. The dotted line
indicates the peak positions of the density disturbance measured with the Gaussian fit.
The dot-dashed line has a slope equal to the sound speed ($C_s$). (d) Spatial profiles of
velocity (dashed line) and perturbed density (solid line) along the loop at t=240 $\tau_A$. 
The pre-pulse velocity profile (at t=159 $\tau_A$) is overplotted with the dotted line.
(e) Spatial profiles of the wave propagation speed derived from the theoretical prediction
(dotted line) and from the measured path of density disturbance ({\it Diamonds}).
The dashed line represents the velocity for background flow ($V_{bg}(s)$,
taken at t=159 $\tau_A$).  The dot-dashed line indicates the wave speed estimated with 
$C_s+V_{bg}(s)$. The solid line indicates the velocity variation along the flow path.}
\end{figure*}

\begin{figure*}
\epsscale{1.0}
\plotone{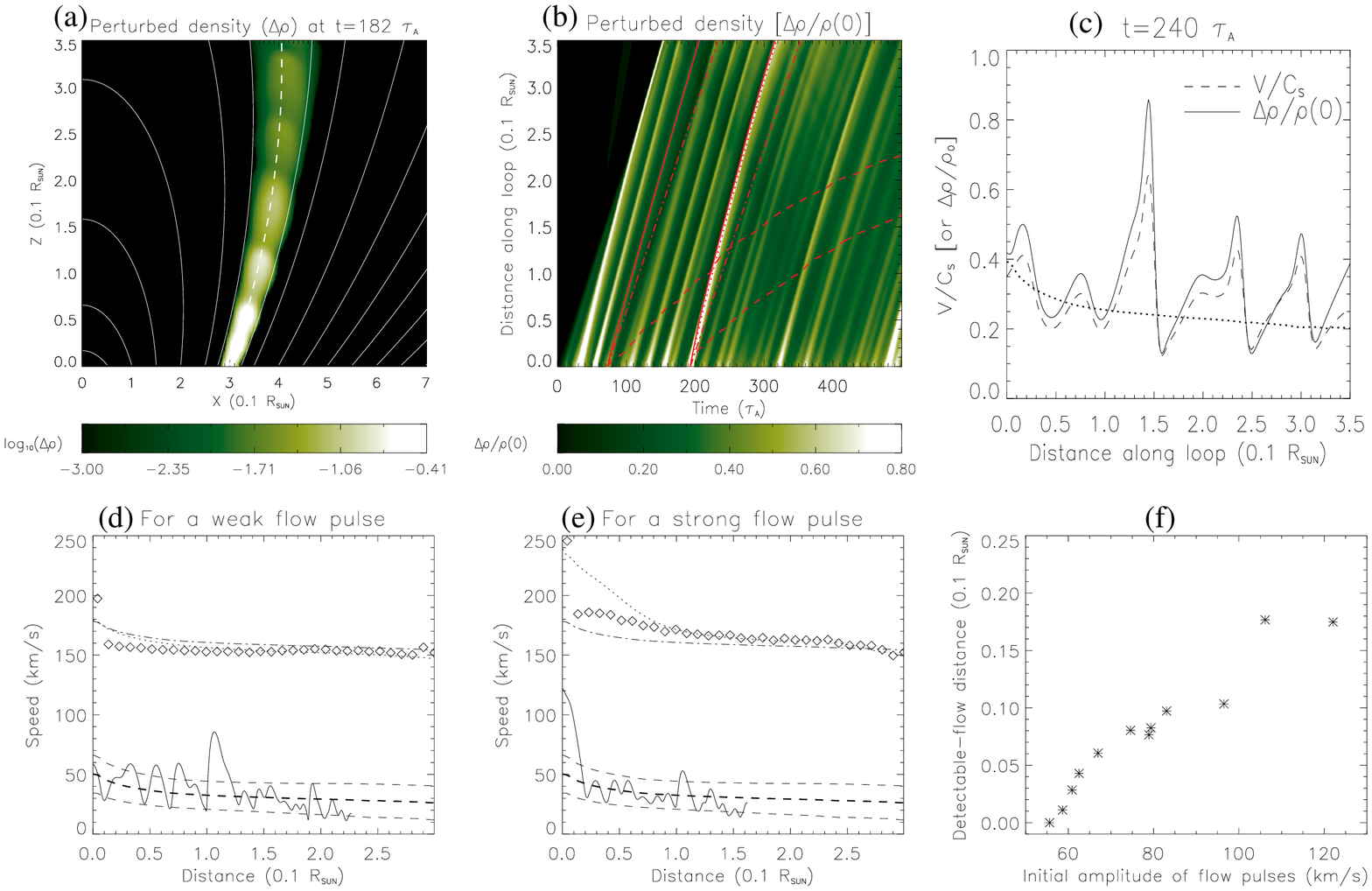}
\caption{ \label{fgbrd} Similar as Figure~\ref{fgsng} but for a broadband upflow driver. 
 The dotted line in (c) indicates the spatial profile of velocity averaged over time.
The annotations for (d) and (e) are same as Figure~\ref{fgsng}(e) but for the two flow
pulses analyzed in (b). The dashed lines are the time-averaged velocity (for background flow)
and the standard deviations.  (f) The detectable distance of flow pulses as a function of 
the initial amplitude (see text for detail).}
\end{figure*}

\section{Numerical results}

First, we report the results in the single pulse case, where an upflow pulse could be produced 
by a small flare. Figures~\ref{fgsng}(a) and (b) show
the snapshots of density and its perturbation in the {\it xz}-plane at $t=240 \tau_A$. The
steady inflow first forms a steady fan-like loop, and then the upflow pulse leads to a density
disturbance traveling along the loop. Figure~\ref{fgsng}(d) shows the spatial profiles of velocity 
and perturbed density ($(\rho(s,t)-\rho(s,0))/\rho(s,0)$) at $t=240 \tau_A$, indicating 
an in-phase relationship between them and a deceleration of background flow at lower heights. 
We calculated the flow and wave paths for this pulse in the time-distance diagram 
(Figure~\ref{fgsng}(c)). Obviously,
the wave path (solid line) coincides well with that of the propagating density
disturbance (PD, dotted line), which is measured by the Gaussian fits to the time profiles 
at each height. In Figure~\ref{fgsng}(e), we compared the measured speeds of the PD
({\it Diamonds}, by taking the derivatives of the time profile of peak positions)
with the wave speeds theoretically predicted ($V(s,t)+C_t(s)$, dotted line) and the velocity of 
injected flow pulse (solid line). We also calculated another estimate of the wave speed using
$C_s+V_{bg}(s)$ (dot-dashed line), where $V_{bg}(s)$ is the velocity for background flow.
We find that the variation of the measured PD speeds agrees well with the theoretically
expected phase speed in the presence of background flow, while the velocity of injected 
flow pulse is much smaller and shows a drastic drop to the background level. This indicates 
that the PD generated in the simulation is dominated by the wave signature,
whereas the contribution of injected high-speed flows is only present at lower heights.

We now present the results of the broadband flow case, where quasi-periodic 
pulses are assumed to be produced by recurrent nanoflares. Figures~\ref{fgbrd}(a) 
show the propagation of quasi-periodic density disturbances in the {\it xz}-plane 
at $t=182 \tau_A$. Figure~\ref{fgbrd}(c) compares the density perturbation and velocity 
profiles along the loop at $t=240 \tau_A$, showing that their variations are roughly 
in phase, consistent with the features of propagating slow magnetosonic waves. 
The background flow along the loop was calculated by averaging the velocity over 
time at each height, showing a decrease with height. 
For a comparison we calculated the wave and flow paths for the two flow pulses, a weak one
(at $t= 75\tau_A$) with the amplitude close to that of background flow and a strong one 
(at $t=195 \tau_A$) with the amplitude close to the sound speed. Figure~\ref{fgbrd}(b) 
shows that for either weak or strong flow pulse, the wave path coincides well with the PD.
For the weak pulse the flow is immediately separated from the PD after injection, suggesting
that the PD is mainly the wave signature, whereas for the strong pulse  
the flow is visible within the PD over a distance of about $D_{flow}$=0.22, where 
$D_{flow}$ (named the detectable-flow distance) is defined as the distance over
which the peak velocity of pulse remains at least 15\% above the background flow. 
Figure~\ref{fgbrd}(f) shows that $D_{flow}$ is proportional to the initial amplitudes of pulses.
Figure~\ref{fgbrd}(e) shows that the propagation speed of the PD generated by this strong pulse
agrees well the predicted wave speed except at the lower heights, where  we notice that the 
speeds of PD are better consistent with the wave speeds estimated by 
$C_s+V_{bg}(s)$. This may be because the flows highly deviate from the
(quasi-) steady state at lower heights, as indicated by a quick deceleration.  
This suggests that the PDs are dominated
by the wave signature with the flow contribution only at lower heights. 
For the 12 generated PDs, we measured their average propagating speeds over
a distance of $s=1$ by the linear fitting to their propagation paths, and obtained
the mean value of 165$\pm$9 km~s$^{-1}$. In comparison, we also measured the average speeds
of their injected flows using the same method, and obtained the mean of 37$\pm$5 km~s$^{-1}$. 
Thus, the density disturbances produced by the upflow pulses propagate at 
a speed on average about four times faster than the corresponding injected flows.

It should be pointed out that the PDs can be modeled in 1D in certain simplified restrictive 
cases. However, as we find from several 1D test calculations, the results may differ 
significantly from the 3D MHD model with more realistic divergent and curved magnetic geometry, 
and with non-uniform injected flow profile. We have also made two additional runs 
for a flux tube with larger and smaller divergences respectively, by adjusting the depth 
of the dipole source, and found similar results in all cases, i.e., the waves dominate PDs 
from very low height. In the case of smaller divergence the height of detectable flow is 
somewhat higher than in the case of larger divergence.

\begin{deluxetable}{lll}
 \tabletypesize{\scriptsize}
 \tablecaption{List of physical parameters used in the model.
\label{tabpar}}
 \tablewidth{0pt}
 \tablehead{\colhead{Quantities} & \colhead{Single pulse case} & \colhead{Broad band case}}
\startdata
\multicolumn{2}{l}{Parameters for flow driver} & \\
\hline
Background flow (km~s$^{-1}$) & 59 & 55 $\pm$ 18 \\
Pulse amplitude (km~s$^{-1}$) & 59 & 10 $-$ 110 \\
Pulse period (min) & 6 & 3 \\
\hline
\multicolumn{2}{l}{Measured quantitie} & \\ 
\hline
Wave speed (km~s$^{-1}$) & 184 & 165 $\pm$ 9 \\
Flow speed (km~s$^{-1}$) & 41 & 37 $\pm$ 5\\
\hline
\multicolumn{2}{l}{Model nomalization parameters} & \\ 
\hline
\multicolumn{2}{l}{Length scale ($a$)} & 0.1 R$_s$ \\
\multicolumn{2}{l}{Magnetic field ($B_0$)} & 100 G \\
\multicolumn{2}{l}{Temperature ($T_0$)} & 1 MK \\
\multicolumn{2}{l}{Number density ($n_0$)} & 1.38$\times{10}^9$ cm$^{-3}$ \\
\multicolumn{2}{l}{Alfv\'{e}n speed ($V_{A0}$)} & 5872 km~s$^{-1}$ \\
\multicolumn{2}{l}{Alfv\'{e}n time ($\tau_A$)} & 12 s\\
\multicolumn{2}{l}{Sound speed ($C_s$)} & 128  km~s$^{-1}$ \\
\multicolumn{2}{l}{Gravitational scale height ($H_0$)} & 60 Mm\\ 
 \enddata
\end{deluxetable}

\section{Discussion and conclusions}
To determine whether propagating intensity disturbances seen in EUV imaging observations
of coronal loops  are flows or waves, we develop a 3D MHD model  
with the geometry and initial state in qualitative agreement with typical
observations. We first use the driver of a single upflow pulse with steady background at 
the footpoints of open-like magnetic fields (i.e., a long loop initially in hydrostatic 
equilibrium that closes outside the 
computational domain) to study the effects of high-speed jets in the persistent upflow 
region observed by Hinode/EIS \citep[e.g.][]{uga11, nis11}. The simulations show that 
the injection of a velocity pulse (with the parameters typically for observations) 
inevitably excites a slow magnetosonic wave disturbance propagating along the loop at the 
phase speed of about the sound speed plus the background flow velocity. This result suggests 
that the observed quasi-periodic PDs,
when interpreted as slow magnetosonic waves, are not necessarily excited by a quasi-periodic 
driver such as the photospheric $p$-mode leakage, but can be produced by recurrent 
small-scale impulsive energy release such as nanoflares \citep{harh08, ofm12}, or 
reconnection jets \citep[e.g.][]{yok95,gon09}. Following this idea, we construct a broadband 
flow driver with repetitive tiny pulses, and successfully 
reproduce the quasi-periodic PDs which are qualitatively consistent with the observations. 
We find whether upflow pulses are small (with amplitudes close to the persistent 
background flow of the order of 20$-$30 of km~s$^{-1}$), or strong (with amplitudes
of the order of the sound speed), the generated PDs in our simulations are dominated by 
the wave signature as their propagation speeds are consistent with the wave speed in the 
presence of flows; for the latter case the injected flows may be detectable at heights up to
$\sim$20 Mm, comparable to the detection lengths of observed PDs \citep{mce06}, but these
flows decelerate rapidly with height primarily due to gravity and the downward  thermal 
pressure gradient that results from the relative density increase due to the initial upflow 
material. 

Our simulations help solve several controversies: the results suggest that persistent upflows 
observed with Hinode/EIS (see Figure~\ref{fgobs}(b)) may be 
a collective effect of unresolved tiny velocity pulses produced by nanoflares at the coronal base, 
while the individual quasi-periodic PDs imaged with TRACE or SDO/AIA (see Figure~\ref{fgobs}(c)) 
correspond to low frequency larger events. This scenario can eliminates the difficulty explaining 
the origin of longer-period (10$-$30 minutes) oscillations (or harmonics) of PDs by the leakage
of global $p$-modes. Note that it is possible that the nanoflare-produced broadband flows are 
modulated by the 5-minutes $p$-mode oscillations in non-sunspot loops \citep{mce06}. 
In addition, if the broadband periodicity of PDs is the effect of nanoflares, they may be used 
to diagnose the energy distribution of nanoflares.  

Our results also suggest a possible connection between the wave and flow 
interpretations for PDs. On the one hand, spectroscopic features such as line blue asymmetries and 
width broadening infer the presence of high-speed ($\sim$ 100 km~s$^{-1}$) outflows 
\citep{dep10, tian11a}. On the other hand, our MHD modeling indicates that any subsonic 
flow pulses injected at the loop footpoints will inevitably excite slow magnetosonic wave 
disturbances propagating ahead of the injected flows. Thus, the flows and waves may both contribute 
to the formatin of PDs. Observations show that the PDs sometimes can reach higher altitudes 
(on the order of the gravitational scale height) at almost constant speed \citep{wan09, mar09, kri12}. 
This feature is consistent with our simulation results and suggests the dominant contribution of waves, 
while high-speed dynamic flows may be only present at lower heights. EIS observations show that the 
line blue asymmetries are detectable mainly within a distance of about 20 Mm above loop's footpoints 
\citep{nis11, tian11b, mci12}.  A future work that compares synthetic line profiles in the PDs with
observations will help test the model and confirm the above scenario.

Although our model is limited to the idealization of an isothermal plasma, it can still model the 
flows produced by impulsive heating.  Because the impulsive heating increases the
thermal pressure to drive the flow along the magnetic field, it does not matter whether the pressure 
pulse is produced by a pulse of density, temperature, or both. In addition, to obtained flow pulses 
(rather than waves) at nearly constant sonic or supersonic speeds as observed for the PDs at higher 
heights, some non-thermal effects such as magnetic slingshot effect due to magnetic reconnection 
\citep[e.g.][]{gon09}, or non-isothermal effects such as the temperature of outflows increasing 
with height \citep{ima11} may be required. However, no evidence for these effects have been found 
in imaging or spectral observations of PDs so far. This also justifies the use of the isothermal 
energy equation in our model. Although the model shows qualitatively that 
PD signatures are dominated by slow magnetosonic waves from low heights using idealized 
active region model, in real coronal loops the quantitative sensitivity of these signatures 
to the details of the field configuration needs to be studied, and any direct comparison 
of the observed signatures to the modeled signatures has to include a proper model of 
the corresponding particular solar magnetic fields.

\acknowledgments
The work of TW was supported by NASA grant NNX12AB34G and the NASA Cooperative Agreement 
NNG11PL10A to CUA. LO acknowledges support by NASA grants NNX12AB34G and NNX11A0686G.

\end{document}